\documentclass{fcs}
\usepackage{bm}

\usepackage{graphicx}
\usepackage[misc]{ifsym}
\usepackage[justification=centering]{caption}
\usepackage{graphicx}
\usepackage{graphics}
\usepackage{booktabs}
\usepackage{subfigure}
\usepackage{algorithm}  
\usepackage{algorithmic} 

\usepackage{amsfonts}
\usepackage{amsmath}                
\usepackage{amsbsy} 
\usepackage{amssymb}  

\usepackage{bigstrut}
\usepackage{multirow}
\usepackage{multicol}
\usepackage[titletoc]{appendix}

\usepackage{geometry}
\renewcommand{\xff}{\raisebox{-.05cm}{\includegraphics{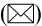}}}
\volumn{ }
\doi{}
\articletype{RESEARCH~ARTICLE}
\copynote{{\copyright} Higher Education Press and Springer-Verlag Berlin Heidelberg 2012}
\ratime{Received month dd, yyyy; accepted month dd, yyyy}
\email{fzhe@whu.edu.cn}
\title{A Correlative Denoising Autoencoder to Model Social Influence for Top-N Recommender System}
\author{Yiteng Pan$^{1}$, \xff Fazhi He $^{1}$ and Haiping Yu $^{1}$}
\address{{1\quad School of Computer Science, Wuhan University, Wuhan 430072, China}}

\def\et{{et al. }}

\begin{document}
	\maketitle
	\setcounter{page}{1}
	\setlength{\baselineskip}{14pt}
	
	\begin{abstract}
		
		In recent years, there are numerous works been proposed to leverage the techniques of deep learning to improve social-aware recommendation performance. In most cases, it requires a larger number of data to train a robust deep learning model, which contains a lot of parameters to fit training data. However, both data of user ratings and social networks are facing critical sparse problem, which makes it not easy to train a robust deep neural network model. Towards this problem, we propose a novel \textit{Correlative Denoising Autoencoder} (CoDAE) method by taking correlations between users with multiple roles into account to learn robust representations from sparse inputs of ratings and social networks for recommendation. We develop the CoDAE model by utilizing three separated autoencoders to learn user features with roles of rater, truster and trustee, respectively. Especially, on account of that each input unit of user vectors with roles of truster and trustee is corresponding to a particular user, we propose to utilize shared parameters to learn common information of the units that corresponding to same users. Moreover, we propose a related regularization term to learn correlations between user features that learnt by the three subnetworks of CoDAE model. We further conduct a series of experiments to evaluate the proposed method on two public datasets for Top-N recommendation task. The experimental results demonstrate that the proposed model outperforms state-of-the-art algorithms on rank-sensitive metrics of MAP and NDCG.
		
	\end{abstract}
	
	
	\section{Introduction}
	
	With the rapid development of social media in recent years, the massive amount data of social network offers researchers a great opportunity to study user behavior patterns and model user preferences for a variety of applications~\cite{1,2}. These social-aware methods have been widely used for a lot of web services, such as Ciao.com, Epinions.com and Facebook.com. To address the sparse problem of recommender system, there are numerous social-aware recommendation algorithms been developed to help model user preferences by integrating social influence for users.
	
	To model social influence from social network, there are numerous works been proposed to make use of social network to improve recommendation performance~\cite{3,4,5,6,7}. These methods can roughly classify these methods into two categories: single-role based methods~\cite{3,4,5} and multi-role based methods~\cite{8,7}. The single-role based methods model each user with a single role for all user feedbacks, while multi-role methods model each user with different roles in different cases. Intuitively, the multi-role methods are closer to our real life since each user plays different roles in different situations. Experimental results in~\cite{7} and~\cite{8} also prove that multi-role based methods are more accurate than those single-role based methods~\cite{3,4,5}. However, the data of social network are facing critical sparse problem~\cite{9}, which make it not easy to extract exact social information from these data. 
	
	In recent years, deep learning models have achieved a great success~\cite{10,11,12,13,14} besides classical methods~\cite{15,16,17,18,19,20,21,22}. Those non-linear neural networks can automatically learn effective representations from huge amount of data and significantly improve prediction accuracy. In recent years, how to utilize deep learning model to improve recommendation performance has become a hot topic~\cite{23,24,25,26,27,28}. To overcome the data spare problem for recommendation, most researchers propose to utilize the Denoising Autoencoder (DAE)~\cite{29} to learn compact representations from spare data for recommender systems. Some DAE-based methods~\cite{23,27} focus on how to learn user preferences from users' rating information by DAE network directly. Their experimental results demonstrate great improvement while compared to traditional linear models, such as matrix factorization method~\cite{30}. Meanwhile, some other works~\cite{31,25,26,28} attempt to use DAE model to learn compact representations from auxiliary information to help improve CF-based method against data sparse problem, such as content~\cite{31}, tags~\cite{25} or images~\cite{26}.
	
	To make use of social media, there are numbers of works been proposed to utilize the technique of neural network to improve the performance of social recommendation~\cite{28,32,33}. In~\cite{28}, Deng \et propose to learn user representations from ratings by Denoising Autoencoder. Then they propose a new collaborative filtering method by utilizing the learnt results to initialize the preferences of users and measure trust similarities. In~\cite{32}, Guo \et propose an embedding-based recommendation method to exploit the deep structure in social networks and rating patterns. In~\cite{33}, Wang \et propose to learn user representations by deep neural networks with inferred ratings and social networks. They further propose to inject original and inferred social information into the input and hidden layer of neural network to improve recommendation. 
	
	There is a key problem remained open for the task of utilizing deep learning techniques for social recommendation. Although deep neural networks show great learning power for many pattern recognition tasks, it requires a large amount of data for training to keep aware from overfitting in most cases~\cite{10}. However, for that both data of user ratings and social networks are facing critical sparse problem, it is not easy to learn robust representations by deep learning models. Therefore, it is necessary to develop a new model to mine more information from social networks by deep learning models for recommendation.
	
	In this paper, we develop a novel deep learning model of \textit{Correlative Denoising Autoencoder} (CoDAE) to learn user features and take the correlations among user features with multiple roles into account. First, we utilize three separated autoencoders to learn user features independently from heterogeneous information of users with multiple roles of rater, truster and trustee, respectively. Second, we propose to inject a shared parameter matrix into the model to learn common information of users with multiple roles. Third, we propose a related regularization term to build relations for user features with different roles.
	
	Especially, in the view of input layers, each input unit of users with roles of truster and trustee is corresponding to a particular user. Therefore, there may exist some implicit common features between the training parameters that corresponding to same users in input layers for roles of truster and trustee. In CoDAE model, we propose to utilize a shared weight matrix to learn implicit common features for users. Moreover, in the view of middle layers, the output vectors of each user with roles of truster and trustee reflect user's preferences in different perspectives. So that these output vectors for a particular user are similar with each other. This motivates us to propose the related regularization term to build relations among user vectors. With these two methods to mine correlations for each user with multiple roles, the CoDAE model is formulated to simultaneously learn user features from both user ratings and social network by neural networks.

	The main contributions in this paper are summarized as follows:
	
	\begin{itemize}
		\item To learn user features from ratings and social network with different distributions, we propose a novel Correlative Denoising Autoencoder (CoDAE) model by utilizing three separate autoencoders to model users with roles of rater, truster and trustee, respectively. 
		
		\item To address the sparse problem of social network, we propose to utilize a shared weight matrix to learn implicit common information between input units that corresponding to same users.
		
		\item To simultaneously exchange information between user features with multiple roles, we further propose a related regularization term to build relations between hidden layers of the three separated autoencoders.
		
		\item We conduct comprehensive experiments to evaluate the proposed method CoDAE for Top-N recommendation task. Experimental results on two real-world public datasets demonstrate that the CoDAE model significantly outperforms other state-of-the-art algorithms.
	\end{itemize}
	
	\section{Related Work}
	In this section, we discuss the relations of the proposed CoDAE model for recommendation in two perspectives: Trust-aware recommendation and Deep learning for recommendation.

	\subsection{Social recommendation}
	
	With the rapid development of social media, there are numerous works been proposed to leverage social influence to improve web applications~\cite{4,34}. This raises the problem of that how to integrate social influence into recommender systems and attracts a lot of attention in recent years~\cite{3,4,5,6,7}. These social recommendation algorithms have been proved to be effective to address the data sparse problem of conventional CF-based methods. 
	
	Especially, Ma \et~\cite{3} propose the SoRec model to integrate trust relationships by factorized the social matrix, which shared common latent user preferences with CF method. Jamali \et~\cite{4} propose a novel SocialMF model with trust propagations between users. The basic idea is that a user's taste is close to the average preference of his/her trust friends. Ma \et~\cite{5} further propose to take trust strength into account, and then build the SoReg model by incorporating individual social regularization. 
	
	The above algorithms model users with a common shared space for rating data and trust relationships. These single-role based methods do not consider that each user plays different roles in different situations. Therefore, some multi-role methods are proposed to address this problem. For example, Yang \et~\cite{6} propose the TrustMF method to model users with roles of truster and trustee. Then they propose to make predictions by integrating these two kinds of information. In~\cite{7}, Yao \et further propose to take implicit correlation between users whole are similar but not socially connected by modeling users with multi-role.
	
	However, the trust relationships in social network are also facing data sparse problem in most cases~\cite{9}, which may limit the performance of social-aware applications. To overcome this problem for recommender system, we propose to utilize the recent deep learning techniques to learn social influence for each user through denoising autoencoder network, which is quite suitable for sparse data~\cite{29}. 
	
	\subsection{Deep learning for recommendation}
	
	Deep neural network has already been proved to be a powerful learning model for many pattern recognition tasks, such as image recognition~\cite{11}, object detection or neural machine translation. Due to the great learning ability in various domains, it attracts a lot of attention to utilize deep learning techniques for recommendation in recent years.
	
	Especially, the AutoRec~\cite{23} model is proposed by utilizing Denoising Autoencoder to direct learn user feature from rating data. Their results demonstrate that using deep learning techniques has great potential to improve recommendation performance. In~\cite{27}, Wu \et propose to inject user-special vectors in hidden layer of Autoencoder network to exact learn user preferences for top-\textit{n} recommendation. Obviously, these methods are plagued by sparse problem of rating data, which have an adverse impact on recommendation performance. In~\cite{35}, the ECAE model is proposed to learn implicit information from generated labels of users. Their experimental results demonstrate that this is a potential approach to address sparse problem for recommendation. 
	
	Incorporating auxiliary information with rating data is one of the most effective ways to address data sparse problem~\cite{31,26}. In~\cite{31}, Wang \et propose the CDL model, which utilizes Denoising Autoencoder to learn compact representations from content information. These representations are then tightly integrated with matrix factorization to model user preferences for recommendation. The VBPR model~\cite{26} make use of visual features learnt by Convolutional Neural Network (CNN) to help improve Bayesian Personalized Ranking (BPR)~\cite{36} model for top-\textit{n} recommendation. Their theoretical and experimental results demonstrate that it is a valuable idea to introduce auxiliary information for deep learning based recommendation.
	
	However, most existing methods focus on learning compact representations from only one kind of data, i.e., rating data or auxiliary information. As a matter of fact, using neural networks for either kind of data can effectively improve the recommendation performance. It raises an interesting question: how to utilize neural networks to learn representations for multiple kinds of data and fuse them into a unified framework for recommendation? 
	
	Towards this problem, we propose the CoDAE method to learn user features from data of ratings and social network by deep neural networks. Especially, to overcome the data sparse problem of ratings and social network, we propose to build relations between user features with different perspectives in two aspects: training parameters for input layer and output units of middle layer. In this way, we obtain a robust social recommendation model based on deep learning techniques. 
	
	\subsection{Deep learning for social recommendation}
	Motivated by the success of deep learning, there are numerous works been proposed to leverage deep learning techniques to improve social recommendation performance~\cite{28,32,33}. Especially, Deng \et propose the DLMF model~\cite{28} to employ Denoising Autoencoder to initialize latent user features, which used in matrix factorization model. In~\cite{32}, Guo \et propose an embedding-based method to learn user representations in social networks and rating patterns for recommendation. In~\cite{33}, Wang \et propose to learn user features by making use of inferred ratings and social networks for deep neural network.
	
	However, since the data of social network are quite sparse, it is not easy to learn robust representations from social network by deep neural networks, which require a larger amount of data to train a robust model in most cases. Therefore, it is necessary to mine more information from social network to learn robust user features. 
	
	In this paper, we propose a novel structure of deep neural network to learn user features from ratings and social networks. Especially, we propose to mine correlations between user features by shared parameters and related regularization. Experimental results demonstrate that these two methods are quite effective to improve recommendation.
	
	\subsection{Heterogeneous network based recommendation}
	In~\cite{37}, Zhang \et propose a new Joint Representation Learning (JRL) framework to learn user representations from multiple heterogeneous data of reviews, images and ratings. They propose to learn representations from difference sources by separated neural networks. These representations are then jointly integrated to represent users and used to train for top-\textit{n} recommendation. To make this framework flexible to be easily extended for new information sources and avoid re-training in practice, this model doesn't take the advantage of the power of multi-view machine learning.
	
	However, as a matter of fact, the data of social network are very sparse in most case. This problem makes it hard to learn robust representations from social network by a neural network, which requires a larger amount of data to prevent overfitting. Therefore, it is necessary to take the correlations among these representations to improve the accuracy and performance.
	
	In this paper, we propose a novel Correlative Denoising Autoencoder (CoDAE) model to learn robust user features from multiple heterogeneous data of ratings and social network. To achieve an unbiased and efficient prediction function, we utilize three autoenders to learn user features by modeling users with roles of rater, truster and trustee. Then we propose to build relations between these subnetworks in two aspects. First, we utilize a shared training weight matrix in them to automatically learn implicit common information for users. Second, we propose a related regularization term to bridge relations between hidden representations of these subnetworks. 
	
	\section{Proposed method}
	In this section, we introduce the proposed Correlative Denoising Autoencoder (CoDAE), which learns compact and robust representations from rating and trust data. 
	
	\subsection{Problem Description}
	With a set of users $ \mathcal{U}=\{1,...,n\} $ and a set of items $ \mathcal{I}=\{1,...,m\} $, the task of top-\textit{n} recommendation is to recommend a list of N items for each user to meet his/her need. In this model, we have three matrices as input data: rating matrix $ \textbf{R} \in \mathbb{R}^{n \times m} $, truster matrix $ \textbf{T} \in \mathbb{R}^{n \times n} $ and trustee matrix $ \textbf{E} \in \mathbb{R}^{n \times n} $ which is the transposed matrix of $ \textbf{T} $. Since both rating and trust data are very sparse, most entries of these input matrices are unobserved and treated as zeros. In particular, we use $ \textbf{R}_u $, $ \textbf{T}_u $ and $ \textbf{E}_u $ to represent the input vectors of user $ u $ with roles of rater, truster and trustee, respectively. Especially, to model personal interests of users, we use $ \textbf{V}_u \in \mathbb{R}^{k} $ to represent the user-specific vector of user $ u $.

	\subsection{Collaborative Denoising Autoencoder}
	
	In Collaborative Denoising Autoencoder (CDAE) model~\cite{27}, the $ d^{th} $ entry $ \tilde{x}_d $ of input vector $ \textbf{x} $ is randomly overwritten by zero with a probability of $ q $ as following:
	\begin{equation}\label{eq:corr} 
	\begin{aligned}
	P(\tilde{x}_d = \delta x_d)&=1-q \\
	P(\tilde{x}_d = 0) &= \quad q
	\end{aligned}
	\end{equation}
	Where $ \delta = 1/(1-q) $ is used to make the corruption unbiased. 
	
\begin{figure}
		\centering
		\includegraphics[width=6cm]{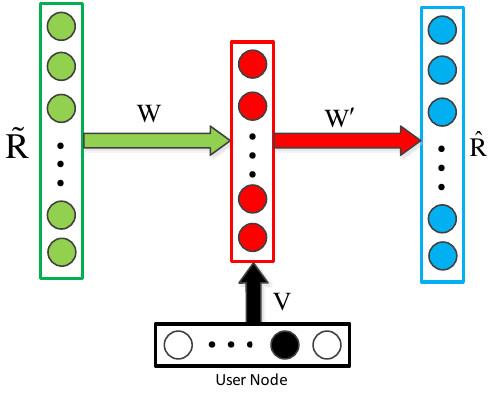}
		\caption{The structure of CDAE model. The user node is used to inject user-specific vector into the hidden layer.}
		\label{fig:CDAE}
	\end{figure}

	As demonstrate in Figure \ref{fig:CDAE}, the CDAE model with one hidden layer is implemented by first mapping the corrupted user feedbacks $ \tilde{R}_u $ into a low-dimensional space through a linear mapping function and then injecting user-specific vectors $ \textbf{V}_u $ into this hidden layer by: 
	\begin{equation}
	\begin{aligned}
	\textbf{H}_u = g(\textbf{W}^T \tilde{\textbf{R}}_u+ \textbf{V}_u +\textbf{b})
	\end{aligned}
	\end{equation}
	Where $ \textbf{W} \in \mathbb{R}^{m \times k} $ is a weight matrix and $ \textbf{b} \in \mathbb{R}^{k} $ is an offset vector. $ g(\cdot) $ is an active function. Note that the dimension of hidden layer is much smaller than that of input layer, i.e., $ k \ll D$.
	
	The user preference $ \textbf{H}_u $ is then mapped back to reconstruct the inputs and predict unobserved entries through another mapping layer by:
	\begin{equation}
	\begin{aligned}	
	\hat{\textbf{R}}_u &= f(\textbf{W}'^T {\textbf{H}}_u+\textbf{b}')\\
	&= f(\textbf{W}'^T (g(\textbf{W}^T \tilde{\textbf{R}}_u+\textbf{V}_u+\textbf{b})) +\textbf{b}')
	\end{aligned}
	\end{equation}
	Where $ \textbf{W}' \in \mathbb{R}^{k \times m} $ and $ \textbf{b} \in \mathbb{R}^{m} $ are the corresponding weight matrix and offset vector to reconstruct inputs through activation function $ f(\cdot) $.
	
	Finally, the objective function of CDAE is formulated by:
	\begin{equation}
	\begin{aligned}	
	\dfrac{1}{2}\sum_{u=1}^{n}\mathit{l}(\hat{\textbf{R}}_u,\textbf{R}_u) + \dfrac{\lambda}{2}(||\textbf{W}||^2_F + ||\textbf{b}||^2_F + ||\textbf{W}'||^2_F + ||\textbf{b}'||^2_F+ ||\textbf{V}||^2_F ) 
	\end{aligned}
	\end{equation}
	Where $ \mathit{l}(\cdot) $ is a loss function to evaluate the reconstruction error for each input vector; $ \lambda $ is a hyper-parameter to control the model complexity to prevent overfitting; $ ||\cdot||_F $ denotes the frobenius norm for regularization term.
	
	To achieve the best performance, we choose the logistic loss as loss function for top-\textit{n} recommendation task in this paper according to the experimental results in~\cite{27}, which is defined by:
	\begin{equation}\label{eq:sp}
	\begin{aligned}
	l(\textbf{y}, \hat {\textbf{y}}) = \sum_{i\in \mathcal{K} \cup \mathcal{S}(t)} -\textbf{y}_{i} log(\hat {\textbf{y}}_{i}) -(\textbf{1}-\textbf{y}_{i})log(\textbf{1}-\hat {\textbf{y}}_{i})
	\end{aligned}
	\end{equation}
	Where $ \mathcal{K} $ denotes the observed entries of vector $ \textbf{y} $ and $ \mathcal{S}(t) $ denotes the entries that sampled from the unobserved entries of vector $ \textbf{y} $, the number of which is $ t $ times of the number of observed entries.
	
	\subsection{Correlative Denoising Autoencoder}
	
	\begin{figure*}
		\footnotesize\centering
		\centerline{\includegraphics[width=10cm]{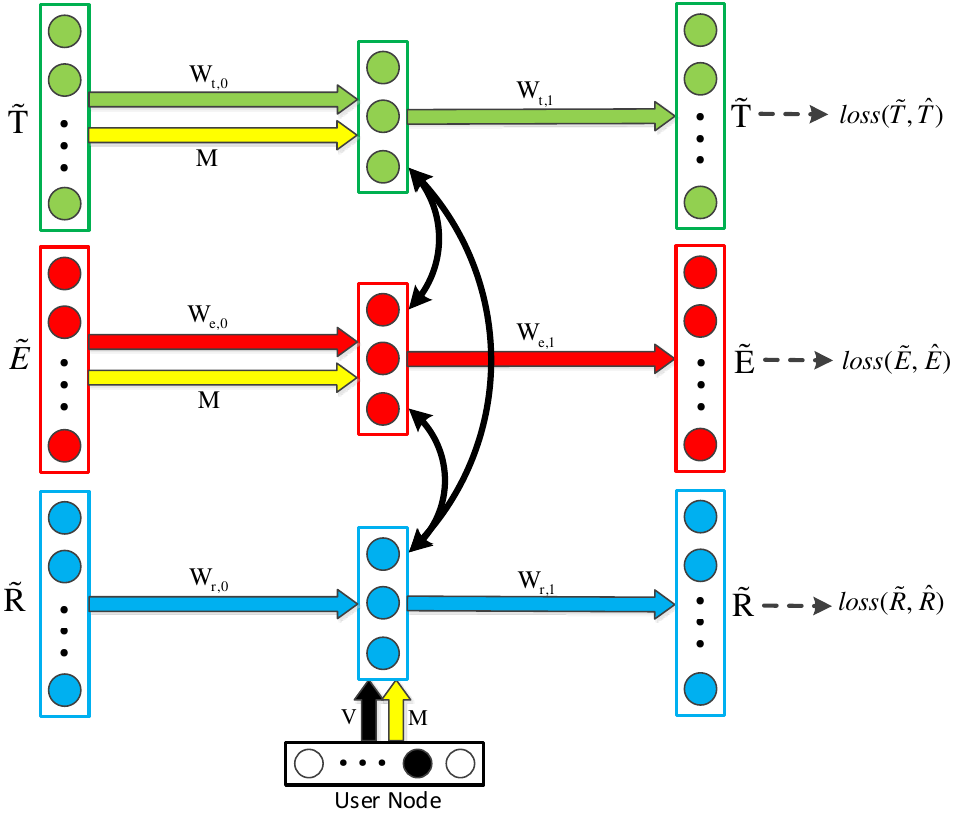}}
		\caption{The structure of CoDAE model which contains three subnetworks to learn compact representations from rating data and social network. The yellow arrows denote the shared weight matrix to learn implicit common information for users. The middle of these subnetworks are crossly connected by a related regularization to build relations for recommendation.}\label{fig:CoDAE}
	\end{figure*}
	
	In recent years, there are a lot of methods been proposed to make use of deep learning techniques to improve social recommendation~\cite{28,32,33}. However, for that there are a lot of parameters to be trained in deep neural networks, it is very easy to fall into overfitting by feeding sparse inputs of user ratings and social networks for these models. Therefore, it is necessary to mine more useful information from these sparse inputs to train deep learning based recommendation models. Towards this problem, we propose a novel deep learning model of Correlative Denoising Autoencoder (CoDAE) to learn user features and take the correlations among user features with multiple roles into account for social recommendation.
	
	As demonstrated in Figure \ref{fig:CoDAE}, to ensure all the information make equal contributions on the prediction results, we utilize three separate subnetworks to independently extract compact representations for each user with roles of rater, truster and trustee. The user nodes for personal biases are then injected into the hidden layer of the subnetwork for raters as that do in CDAE~\cite{27}. Moreover, towards the data sparse problem, we propose to build relations among user features with multiple roles in two aspects: training parameters of input layer and output vectors of middle layer.
	
	Since each unit of the input vectors $ \textbf{T}_u, \textbf{E}_u $ and $ \textbf{V}_u $ is associated with a particular user, we argue that there exists some common information between the training parameters for the input units that associated with the same users. To automatically learn the implicit common information, we denote a shared weight matrix $ \textbf{M} \in \mathbb{R}^{n \times k} $ to represent the common features for these three inputs. In particular, we utilize $ \textbf{M}_u $ to represent $ u^{th} $ vector of $ \textbf{M} $, which is corresponding to the shared common preference of user $ u $.
	
	With the corrupted inputs $\tilde{\textbf{T}}$, $\tilde{\textbf{E}}$ and $\tilde{\textbf{R}} $, which are generated from ${\textbf{T}}$, ${\textbf{E}}$ and ${\textbf{R}} $ through the distribution of Equation~\ref{eq:corr}, the hidden representations $ \textbf{H}_{r,u} \in \mathbb{R}^{k} $, $ \textbf{H}_{t,u} \in \mathbb{R}^{k} $ and $ \textbf{H}_{e,u} \in \mathbb{R}^{k} $ of user $ u $ with roles of rater, truster and trustee are respectively computed by the following mapping functions:
	\begin{equation}\label{eq:features}
	\begin{aligned}	
	\textbf{H}_{r,u} &= g(\alpha\textbf{M}_{u} + (1-\alpha)\textbf{W}^T_{r,0} \tilde{\textbf{R}}_u + \textbf{V}_{u} +  \textbf{b}_{r,0}), \\
	\textbf{H}_{t,u} &= g(\alpha\textbf{M}_{u}^T \tilde{\textbf{T}}_u + (1-\alpha)\textbf{W}^T_{t,0} \tilde{\textbf{T}}_u  + \textbf{b}_{t,0}),\\
	\textbf{H}_{e,u} &= g(\alpha\textbf{M}_{u}^T \tilde{\textbf{E}}_u + (1-\alpha)\textbf{W}^T_{e,0} \tilde{\textbf{E}}_u  + \textbf{b}_{e,0})
	\end{aligned}
	\end{equation}
	Where $ \alpha $ is the hyper-parameter that used to balance the influence of shared weight matrix $ \textbf{M} $. The parameters of $ \{\textbf{W}_{r,0} \in \mathbb{R}^{m\times k}, \textbf{b}_{r,0}\in \mathbb{R}^{k} \} $, $ \{\textbf{W}_{t,0} \in \mathbb{R}^{n\times k}, \textbf{b}_{t,0}\in \mathbb{R}^{k} \} $ and $ \{\textbf{W}_{e,0} \in \mathbb{R}^{n\times k}, \textbf{b}_{e,0}\in \mathbb{R}^{k} \} $ are the weight matrices and offset vectors to map user preferences into a low-dimensional space through active function $ g(\cdot) $. 
	
	The hidden representations $ \textbf{H}_{r,u} $, $ \textbf{H}_{t,u} $ and $ \textbf{H}_{e,u} $ of user $ u $ are then mapped back to predict the clean inputs by: 
	\begin{equation}
	\begin{aligned}	
	\hat{\textbf{R}}_u &= f(\textbf{W}^T_{r,1} \textbf{H}_{r,u} + \textbf{b}_{r,1}),\\
	\hat{\textbf{T}}_u &= f(\textbf{W}^T_{t,1} \textbf{H}_{t,u} + \textbf{b}_{t,1}),\\
	\hat{\textbf{E}}_u &= f(\textbf{W}^T_{e,1} \textbf{H}_{e,u} + \textbf{b}_{e,1})
	\end{aligned}
	\end{equation}
	Where $ \{\textbf{W}_{r,1} \in \mathbb{R}^{m\times k}, \textbf{b}_{r,1}\in \mathbb{R}^{k} \} $, $ \{\textbf{W}_{t,1} \in \mathbb{R}^{n\times k}, \textbf{b}_{t,1}\in \mathbb{R}^{k} \} $ and $ \{\textbf{W}_{e,1} \in \mathbb{R}^{n\times k}, \textbf{b}_{e,1}\in \mathbb{R}^{k} \} $ are the weight matrices and offset vectors of the decoder activation function $ f(\cdot) $ for the three subnetworks in Figure~\ref{fig:CoDAE}.

	\begin{figure}
		\centering
		\subfigure[User-item matrix of rater $ \{a,b\} $ and item $ \{A,B,C,D\} $.]{
			\label{fig:empA} 
			\includegraphics[width=3.5cm]{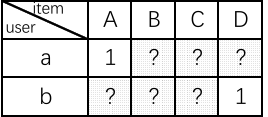}}
		\hspace{0.3in}
		\subfigure[User-user matrix of truster $ \{a,b\} $ and trustee $ \{c,d,e,f\} $.]{
			\label{fig:empB} 
			\includegraphics[width=3.5cm]{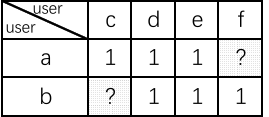}}
		\caption{An example of users in views of ratings and social network.}
		\label{fig:rel} 
	\end{figure}
	
	Although users show different characters when they are playing different roles, the key personalities of each user will not change in most cases. In other words, although the representations for each user with multiple roles are different, there are something in common among these representations for each user. For example, as demonstrated in Figure \ref{fig:rel}, user $ a $ and $ b $ are not similar in Figure \ref{fig:empA}, but they are similar as trusters in Figure \ref{fig:empB}. This may indicate that user preferences of a and b are similar to a certain degree in view of raters for that there are too few observed common ratings to exact model their preferences. Therefore, it is necessary to build relations between these representations with multiple roles to exchange information for sparse inputs of user ratings and social networks.
	
	Especially, we propose a novel regularization term to build relations between user features with multiple roles by minimizing the distances among output vectors of middle layers in CoDAE network as follows:
	\begin{equation}
	\begin{aligned}	
	Rel = \dfrac{\beta}{2}\sum_{u=1}^{n}(||\textbf{H}_{r,u} - \textbf{H}_{t,u}||_F^2 +  ||\textbf{H}_{r,u} - \textbf{H}_{e,u}||_F^2 + ||\textbf{H}_{t,u} - \textbf{H}_{e,u}||_F^2)
	\end{aligned}
	\end{equation}
	Where $ \beta $ is the hyper-parameter to control the importance of related regularization term. A larger value of $ \beta $ indicates that users share more common features with multiple roles. If $ \beta \rightarrow \infty $, the user representations of $ \textbf{H}_{r,u} $ , $ \textbf{H}_{t,u} $ and $ \textbf{H}_{e,u}$ will be optimized to almost equal values by minimizing this related regularization term. If $ \beta = 0 $, this regularization makes no impact on the results of CoDAE model. For that each user show different but similar preferences when playing different roles in real life, the best value of $ \beta $ should be determined by experimental results. 
	
	Formally, we learn the parameters of CoDAE model by minimize the following objective function:
	\begin{equation}\label{eq:CoDAE}
	\begin{aligned}
	\mathcal{L} &= \dfrac{1}{2}\sum_{u=1}^{n} (l({\textbf{R}_u}, \hat{\textbf{R}}_u) + l({\textbf{T}_u}, \hat{\textbf{T}}_u) + l({\textbf{E}_u}, \hat{\textbf{E}}_u) )\\
	& + \dfrac{\beta}{2}( ||\textbf{H}_r - \textbf{H}_t||_F^2 + ||\textbf{H}_r - \textbf{H}_e||_F^2 + ||\textbf{H}_t - \textbf{H}_e||_F^2 )\\ 
	& + \dfrac{\lambda}{2}\sum_{\substack {n \in \{r, t, e\}\\l\in \{0, 1\}}} (|| \textbf{W}_{n,l} ||_F^2 + || \textbf{b}_{n,l} ||_F^2) + \dfrac{\lambda}{2}\sum_{u=1}^{n}(|| \textbf{V}_u ||_F^2+ || \textbf{M}_u ||_F^2)\\
	\end{aligned}
	\end{equation}
	Where $ \lambda $ is the corresponding hyper-parameter to control the model complexity to prevent overfitting; 
	
	Finally, after training the CoDAE model by minimizing loss function of Equation (\ref{eq:CoDAE}), the unobserved entries of $ \textbf{R}_u $ is predicted by:
	\begin{equation}\label{eq:predict}
	\begin{aligned}	
	\hat {\textbf{R}}_u = f(\textbf{W}^T_{r,1} g(\textbf{W}^T_{r,0} {\textbf{R}}_u + \textbf{V}_{u} + \textbf{M}_{u}  + \textbf{b}_{r,0}) + \textbf{b}_{r,1})
	\end{aligned}
	\end{equation}
	Where $ \hat {\textbf{R}}_u $ is the predicted vector of user $ u $. Each $ i^{th} $ unit of $ \hat {\textbf{R}}_u $ stands for the predicted score of user $ u $ for item $ i $. The recommendation item list for user $ u $ is then generated by selecting N items with highest scores from unobserved entries of $ {\textbf{R}}_u $.
	
	\subsection{Model learning}
	The CoDAE model is implemented based on the powerfull deep learning framework of PyTorch, which can automatically compute gradient with strong GPU acceleration. We apply the widely used stochastic gradient descent (SGD) method to learn parameters for CoDAE model. During each iteration, the parameters in CoDAE model are updated as following:
	\begin{equation}
	\begin{aligned}
	{\theta} ^{t+1} =  {\theta} ^{t} - \alpha g_{\theta}^{t}
	\end{aligned}
	\end{equation}
	Where $ \theta^{t} $ represents the value of parameters at iteration $ t $. $ g_{\theta}^{t} $ indicate the corresponding gradients for $ \theta^{t} $. $ \alpha $ is the learning rate during training process.
	
	\subsection{Model complexity analysis}

	As demonstrated in Figure \ref{fig:CoDAE}, the CoDAE model is consist of three subnetworks to learn user features with multiple roles of rater, truster and trustee. For each subnetwork, the input data with dimensionality of $ m $ is first mapped into a middle layer with dimensionality of $ k $ and then mapped back to predict input data. For each iteration, since the CoDAE model is trained over all users, the whole model complexity of CoDAE is $ O(nmk+nnk+nnk) = O(kn(m+n)) $. 
	
	In~\cite{27}, Wu \et propose a novel sampling strategy to reduce the model complexity of neural networks of autoencoder for sparse inputs. For that the CoDAE is also developed based on the structure of autoencoder, this sampling strategy can also be used for CoDAE. Especially, due to the sparse problem of user ratings and social networks, the number of unlabeled data are much larger than that of labeled data. It is unnecessary to compute gradients for all unlabeled data since they contain very little useful information. In this way, the training complexity of CoDAE is $ O(kn(\bar m+\bar n+\bar n')) $, where $ \bar m $, $ \bar n $ and $ \bar n' $ denote the mean numbers of positive labels of users with roles of rater, truster and trustee, respectively. For that the data of user ratings and social networks are very sparse, the values of $ \bar m $, $ \bar n $ and $ \bar n' $ are much smaller than that of $ m $, $ n $ and $ n $, respectively. Therefore, the ECAE model is also computable for big data applications.
	
	\section{Experiments and Results}
	In this section, we demonstrate comprehensive experimental results to evaluate the proposed CoDAE model for top-\textit{n} recommendation task. Our experiments are designed to answer the following questions:
	\begin{itemize}
		\item[-] How does CoDAE model compare with other state-of-the-art related methods?
		\item[-] How does CoDAE model perform for sparse users?
		\item[-] How do the methods of shared parameters and related regularization impact the performance?
		\item[-] How does dropout probability impact on the performance?
		\item[-] How does CoDAE model perform with different dimensions?
	\end{itemize}

	\subsection{Datasets}
	To compare the proposed CoDAE model with state-of-the-arts methods, we use two public real-world datasets for comparison: Ciao and Epinions datasets~\cite{38}. These two datasets contain both rating data and trust relationships, which are crawled from two famous e-commerce websites, i.e., the Ciao.com and Epinions.com. On these websites, users can rate each item with an integer number which range from $ 1 $ to $ 5 $ and build trust relationship with any other users to help making decisions. The trust relationships in social network are formulated in binary format, where $ 1 $ for trust and $ 0 $ for unobserved relationship. The statistics of these two datasets are shown in Table~\ref{tab:sta}. 
	
	\begin{table}[!htbp]
		\footnotesize
		\begin{center}
			\caption{Statistics of Epinions and Ciao}\label{tab:sta}
			\begin{tabular}{|l|c|c|}
				\hline
				Dataset & Ciao  & Epinions \bigstrut\\
				\hline
				Number of users & 7,375 & 22,166 \bigstrut\\
				\hline
				Number of items & 106,797 & 296,277 \bigstrut\\
				\hline
				Number of ratings & 284,086 & 922,267 \bigstrut\\
				\hline
				Number of social links & 111,781 & 300,548 \bigstrut\\
				\hline
				Rating sparsity & 0.036\% & 0.014\% \bigstrut\\
				\hline
				Social sparsity & 0.205\% & 0.061\% \bigstrut\\
				\hline
			\end{tabular}%
			
		\end{center}
	\end{table}
	
	The top-\textit{n} recommendation task is to recommend a personalized list of items for each user to help finding what he/she wants. This task is widely adopted by many existing works~\cite{27}. Since this paper is mainly focus on the top-\textit{n} recommendation task, we remove the rating scores that less than four stars and remain the others with score of one for all datasets~\cite{27}. We then iteratively drop users and items with less than 5 ratings. The statistics of processed datasets are listed in Table \ref{tab:sta2}. 
	
	\begin{table}[!h]
		\footnotesize
		\begin{center}
			\caption{Statistics of the processed datasets}\label{tab:sta2}
			\begin{tabular}{|l|c|c|}
				\hline
				Dataset & Ciao  & Epinions \bigstrut\\
				\hline
				Number of users & 5,072 & 18,096 \bigstrut\\
				\hline
				Number of items & 8,155 & 22,386 \bigstrut\\
				\hline
				Number of ratings & 101,116 & 355,132 \bigstrut\\
				\hline
				Number of social links & 85,916 & 252,101 \bigstrut\\
				\hline
				Rating sparsity & 0.244\% & 0.088\% \bigstrut\\
				\hline
				Social sparsity & 0.334\% & 0.077\% \bigstrut\\
				\hline
			\end{tabular}%

		\end{center}
	\end{table}
	
	\subsection{Evaluation Metrics}
	
	The task of top-\textit{n} recommendation is to recommender a list with N items for each user to fit his/her potential need to maximum level. Relevant studies demonstrate that this task is closer to real world scenario than rating prediction. It means that rating error metrics, such as Root Mean Squared Error (RMSE) and Mean Absolute Error (MAE) are difficult to reflect the performance of recommender systems exactly~\cite{39}. 
	
	Therefore, we choose two ranking-based metrics to evaluate performance in our experiments: Mean Average Precision (MAP@N) and Normalized Discounted Cumulative Gain (NDCG@N). These two metrics take the position influence of predicted list into account and are widely used in many commercial systems~\cite{40}. 
	
	\textbf{Mean Average Precision}. 
	MAP is an improved metric of Precision by considering the performance at all positions of top-\textit{n} item list. The Precision metric is defined by:
	\begin{equation}
	\begin{aligned}	
	P@N&= |I_{u}\cap \hat I_{N,u}|/N\\
	\end{aligned}	
	\end{equation}
	Where $ I_{u} $ represents the items that user $ u $ has rated in the test set; $ \hat I_{N,u} $ indicates the $ N $ items with highest predicted score in unrated item set for user $ u $. Then we have the metric of Average Precision, which is defined as following:
	\begin{equation}
	\begin{aligned}	
	AP@N&= \dfrac{\sum_{k=1}^{N} P@k\times rel(k)}{min\{N,|I_u|\}}
	\end{aligned}	
	\end{equation}
	Where $ rel(k) $ indicate whether the item at rank $ k $ is adopted or not. The Mean Average Precision (MAP@N) is the mean value of AP@N across all users.
	
	\textbf{Normalized Discounted Cumulative Gain}. This metric considers the performance at all positions on the predicted list by giving higher weights to the items with higher predicted scores. Especially, the NDCG is defined as:
	\begin{equation}
	\begin{aligned}	
	NDCG@N&= Z_n\sum_{k=1}^{N}\frac{2^{rel(k)}-1}{log_2(k+1)}
	\end{aligned}	
	\end{equation}
	Where $ Z_n $ is the normalized term over the ideal value iDCG. The mean value of NDCG@N across all users is reported in our experiments.

	\subsection{Comparisons with previous models}
	
	Since we focus on the top-\textit{n} recommendation task, it is unsuitable to compare with the methods designed for rating prediction tasks, such as SVD++~\cite{41} and TrustSVD~\cite{42}. In this section, we compare the proposed CoDAE model with several state-of-the-art top-\textit{n} recommendation algorithms as following:
	
	\begin{itemize}
		\item \textbf{Pop}: This is a popular baseline algorithm which make predictions based on how many people have rated on a particular item.
		
		\item \textbf{BPR}~\cite{36}. This is a ranking based algorithm which is implemented by learning the relations between positive and negative items for each user. 
		
		\item \textbf{GBPR}~\cite{43}. This method relaxes the individual and independence hypothesis in BPR model. The authors propose a new improved model by taking group influence into account. The group size is fixed to 5 as suggested in~\cite{43}.
		
		\item \textbf{SBPR}~\cite{44}. This work proposes to improve the BPR model by considering social connections with the assumption that users tend to assign higher ranks to items that their friends like.
		
		\item \textbf{CDAE}~\cite{27}. The authors utilize the \textit{Denoising Autoencoder} (DAE) technique to improve top-\textit{n} recommendation performance. Especially, they propose to inject user-special vectors into hidden layer of DAE network to further improve recommendation accuracy.
		
		\item \textbf{TDAE}~\cite{33}. This is a novel deep learning based social recommendation method. In this work, the authors propose to learn user representations from both original and inferred data of user ratings and social networks. They also propose to inject social information into the hidden layer of neural network to further improve recommendation.
		
	\end{itemize}
	
	As we focus on learning user preferences with roles of rater, truster and trustee by Denoising Autoencoders, we mainly compared with CF-based and DAE-based methods. We leave out the DLMF method in~\cite{28}, since the performance differences may be result from the community effect which considered in~\cite{28} but not in the proposed CoDAE model. Moreover, for that the TDAE model is developed for rating prediction task, we utilize the logistic loss function and sampling strategy of \ref{eq:sp} that used in CDAE to adjust the TDAE model for item recommendation task.
	
	For all comparison methods, we tune the hyper-parameters carefully by trial and error method in our experiments or according to corresponding reference for each dataset to ensure that each method achieves the best performance for fair comparisons. For each user, we randomly choice $ 80\% $ of the rating data for training process and the others for testing process. We conduct each experiment for 5 times and report the mean performance to fairly compare with other methods. For CDAE and CoDAE methods, we set the drop probability $ q $ to 0.1 and set sample rate $ t $ in Equation \ref{eq:sp} to 5 according to the experimental results in~\cite{27}.

	\begin{table*}[!ht]
		\footnotesize
		\tabcolsep 5pt 
		\begin{center}
			\caption{Comparison results on all users}\label{tab:all}
			\begin{tabular}{|l|l|c|c|c|c|c|c|c|c|}
				\hline
				Dataset & Metrics & Pop   & BPR   & GBPR  & SBPR  & CDAE  & TDAE  & CoDAE & Improve \bigstrut\\
				\hline
				Ciao  & MAP@10 & 0.0210  & 0.0199  & 0.0229  & 0.0219  & 0.0289  & \textbf{0.0299 } & \textbf{0.0306 } & 2.34\% \bigstrut\\
				\cline{2-2}k=10  & NDCG@10 & 0.0369  & 0.0357  & 0.0408  & 0.0402  & 0.0504  & \textbf{0.0521 } & \textbf{0.0528 } & 1.34\% \bigstrut\\
				\hline
				Ciao  & MAP@10 & 0.0210  & 0.0201  & 0.0234  & 0.0240  & 0.0284  & \textbf{0.0303 } & \textbf{0.0311 } & 2.64\% \bigstrut\\
				\cline{2-2}k=20  & NDCG@10 & 0.0369  & 0.0360  & 0.0419  & 0.0435  & 0.0500  & \textbf{0.0533 } & \textbf{0.0541 } & 1.50\% \bigstrut\\
				\hline
				Ciao  & MAP@10 & 0.0210  & 0.0232  & 0.0247  & 0.0255  & 0.0291  & \textbf{0.0320 } & \textbf{0.0329 } & 2.81\% \bigstrut\\
				\cline{2-2}k=100 & NDCG@10 & 0.0369  & 0.0424  & 0.0437  & 0.0456  & 0.0510  & \textbf{0.0549 } & \textbf{0.0561 } & 2.19\% \bigstrut\\
				\hline
				\hline
				Epinions & MAP@10 & 0.0080  & 0.0100  & 0.0115  & 0.0072  & 0.0143  & \textbf{0.0144 } & \textbf{0.0145 } & 0.69\% \bigstrut\\
				\cline{2-2}k=10  & NDCG@10 & 0.0153  & 0.0194  & 0.0218  & 0.0144  & 0.0265  & \textbf{0.0271 } & \textbf{0.0272 } & 0.37\% \bigstrut\\
				\hline
				Epinions & MAP@10 & 0.0080  & 0.0127  & 0.0126  & 0.0090  & 0.0142  & \textbf{0.0149 } & \textbf{0.0151 } & 1.34\% \bigstrut\\
				\cline{2-2}k=20  & NDCG@10 & 0.0153  & 0.0237  & 0.0237  & 0.0173  & 0.0266  & \textbf{0.0277 } & \textbf{0.0280 } & 1.08\% \bigstrut\\
				\hline
				Epinions & MAP@10 & 0.0080  & 0.0163  & 0.0082  & 0.0161  & 0.0164  & \textbf{0.0168 } & \textbf{0.0171 } & 1.79\% \bigstrut\\
				\cline{2-2}k=100 & NDCG@10 & 0.0153  & 0.0291  & 0.0158  & 0.0301  & 0.0304  & \textbf{0.0312 } & \textbf{0.0316 } & 1.28\% \bigstrut\\
				\hline
			\end{tabular}%
			
		\end{center}
	\end{table*}

	\subsubsection{Validations on all users}
	
	The comparison results on all users are demonstrated in Table~\ref{tab:all}. We use two ranking-based metrics to evaluate the recommendation accuracy: MAP@10 and NDCG@10. This table shows the comparison results with dimensions of 10,20 and 100 on two public datasets: Ciao and Epinions.
	
	In Table~\ref{tab:all}, we can see that the CoDAE model performs better than other comparison methods at least of $ 0.69\% $ and $ 0.37\% $ for all datasets on metrics of MAP@10 and NDCG@10, respectively. This fact shows that it is effective to improve recommendation by considering correlations among user features with multiple roles. Moreover, the improvements of CoDAE is more significant while the dimensionality of k growing from 10 to 100. This may indicate that it is easier to learn correlations of user features with multiple roles by utilizing a larger value of dimensionality. In addition, the improvements of CoDAE on Ciao dataset are much more significant than that on Epinions dataset. It may be caused of that the data of Epinions are much sparser than that of Ciao as demonstrated in Table \ref{tab:sta2}, which makes it more difficult to learn correlations of user features by CoDAE model.

	\subsubsection{Validations on sparse users}
	
	\begin{figure}[!htbp]
		\centering
		\includegraphics[width=8cm]{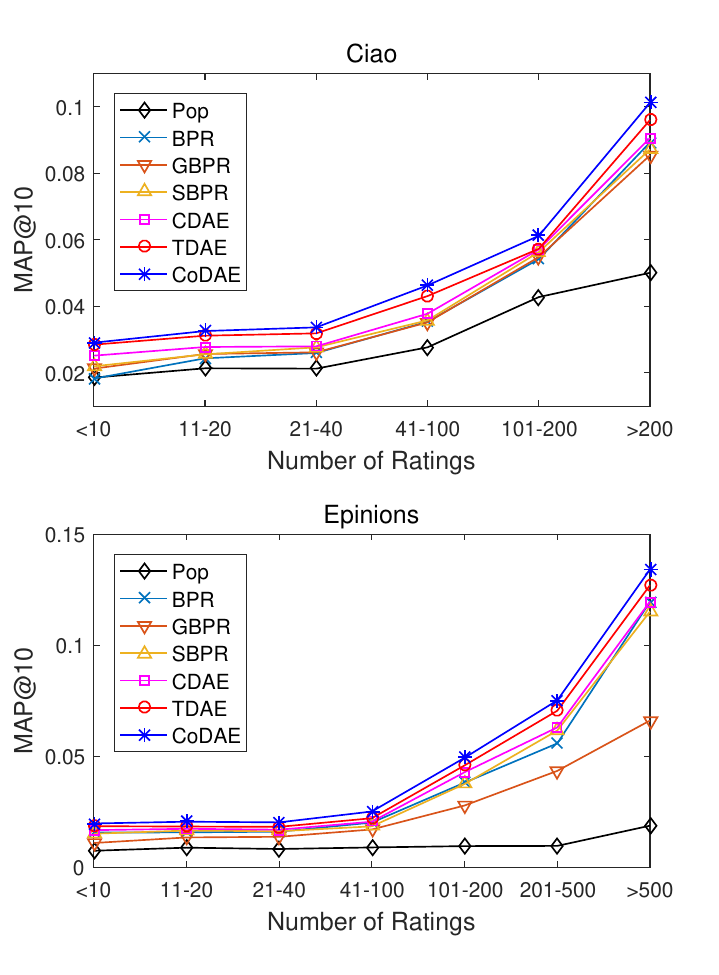}
		\caption{Validations on sparse users with different number of ratings on Ciao and Epinions datasets.}
		\label{fig:Cold}
	\end{figure}
	
	It is well known that the CF-based recommender systems are facing critical data sparse problem, which may degrade the recommendation performance. To validate the performance of CoDAE model in views of sparse users, we demonstrate the performance of users with different rating numbers on metric of MAP@10 with $ k=100 $ in Figure~\ref{fig:Cold}. Since the evaluation results on metric of MAP@10 is consistent with other ranking-based metrics, we omit the performance on other metrics. 
	
	As we can see in Figure \ref{fig:Cold}, the performance of all methods gets better with the number of ratings increases. It indicates that the amount of rating data has a great impact on the recommendation performance and the sparse problem is a critical problem for the recommender systems. Moreover, the CoDAE model performs better than other comparison methods for users with different numbers of ratings. It proves that the CoDAE model is quite effective to address the data sparse problem for recommendation by taking correlations of user features into consideration. In addition, compared with TDAE, the improvements of CoDAE for users with no more than 10 ratings are quite small on both Ciao and Epinions datasets. For that it is difficult to learn robust features from users with too few ratings, it is not easy to learn correlations from these unreliable user features.
	
	\subsection{Impact of shared parameters}
	
	\begin{figure}[!htbp]
		\centering
		\includegraphics[width=8cm]{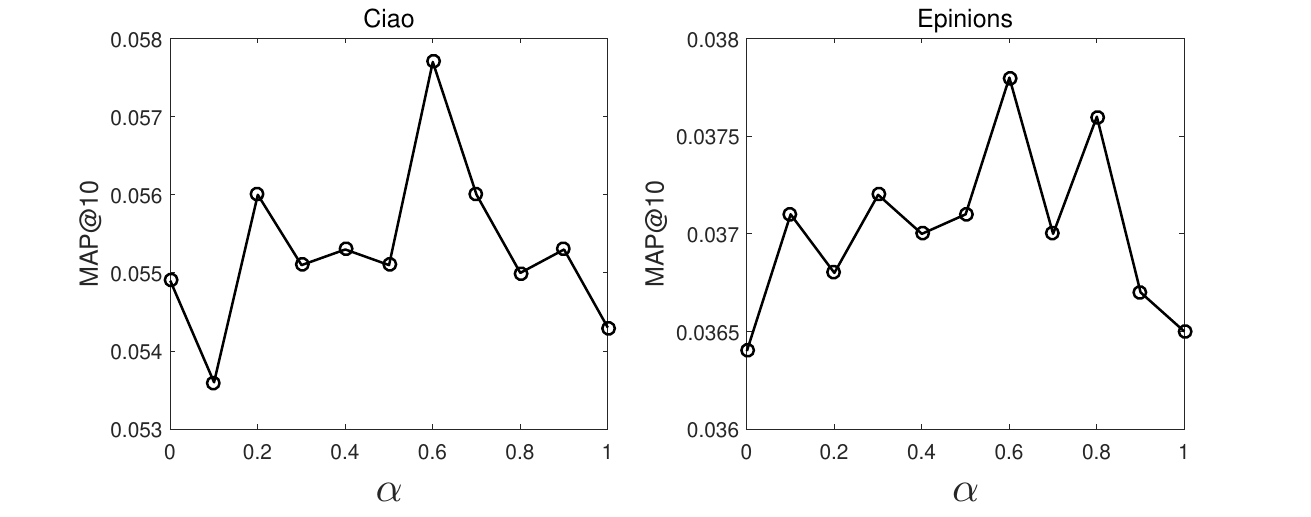}
		\caption{The impact of $ \alpha $ on the performance of CoDAE model.}
		\label{fig:alpha}
	\end{figure}
	
	We conduct a series of experiments to study the influence of shared parameters in Equation (\ref{eq:features}). Especially, $ \alpha $ is used to control the influence of shared parameters $ \textbf{M} $ in CoDAE. A larger value of $ \alpha $ indicates more influence of shared parameters.
	
	In Figure \ref{fig:alpha}, we demonstrate the experimental results of CoDAE model with $ \alpha \in {0,0.1, ... ,1} $. As we can see, the CoDAE model achieves best performance with $ \alpha=0.6 $ in both datasets of Ciao and Epinions. This fact indicates that it is necessary to make use of shared parameters to mine common information of the input vectors for each user with roles of rater, truster and trustee. Moreover, the performance curve of CoDAE in Figure \ref{fig:alpha} is not quite flat. This may be caused by the stochastic training strategy of CoDAE model. Therefore, the best values of $ \alpha $ for different datasets should be determined according to experimental results for each dataset.
	
	\subsection{Impact of related regularization}
	
	\begin{table*}[!htbp]
		\footnotesize
		\begin{center}
			\caption[center]{Impact of related regularization term $ \beta $}\label{tab:ter}
			\begin{tabular}{|l|c|c|c|c|c|c|c|}
				\hline
				$ \beta $    & 0     & 0.001 & 0.005 & 0.01  & 0.05  & 0.1   & 0.2 \bigstrut\\
				\hline
				Ciao  & 0.0461  & 0.0464  & 0.0500  & 0.0514  & 0.0523  & \textbf{0.0547 } & 0.0530  \bigstrut\\
				\hline
				Epinions & 0.0309  & 0.0312  & 0.0313  & \textbf{0.0316 } & 0.0299  & 0.0293  & 0.0276  \bigstrut\\
				\hline
			\end{tabular}%
			
		\end{center}
	\end{table*}
	
	Parameter $ \beta $ are used to control the influence of related regularization. Larger values of $ \beta $ indicate that there exist more correlations between user features of multiple roles. We conduct a group of experiments to evaluate the influence of related regularization term and report the results in Table~\ref{tab:ter}. We perform these experiments with $ k=100 $. 
	
	In this table, we can see that the performance of CoDAE increases with the value of $ \beta $ grows and up to a point. When the value of $ \beta $ become even larger, the performance decreases due to overfitting. These experimental results indicate that the social influence makes a great impact on the recommendation accuracy and the value of $ \beta $ should be adjusted for different datasets.
	
	\subsection{Impact of dropout probability}

	\begin{figure}[!htbp]
		\centering
		\includegraphics[width=8cm]{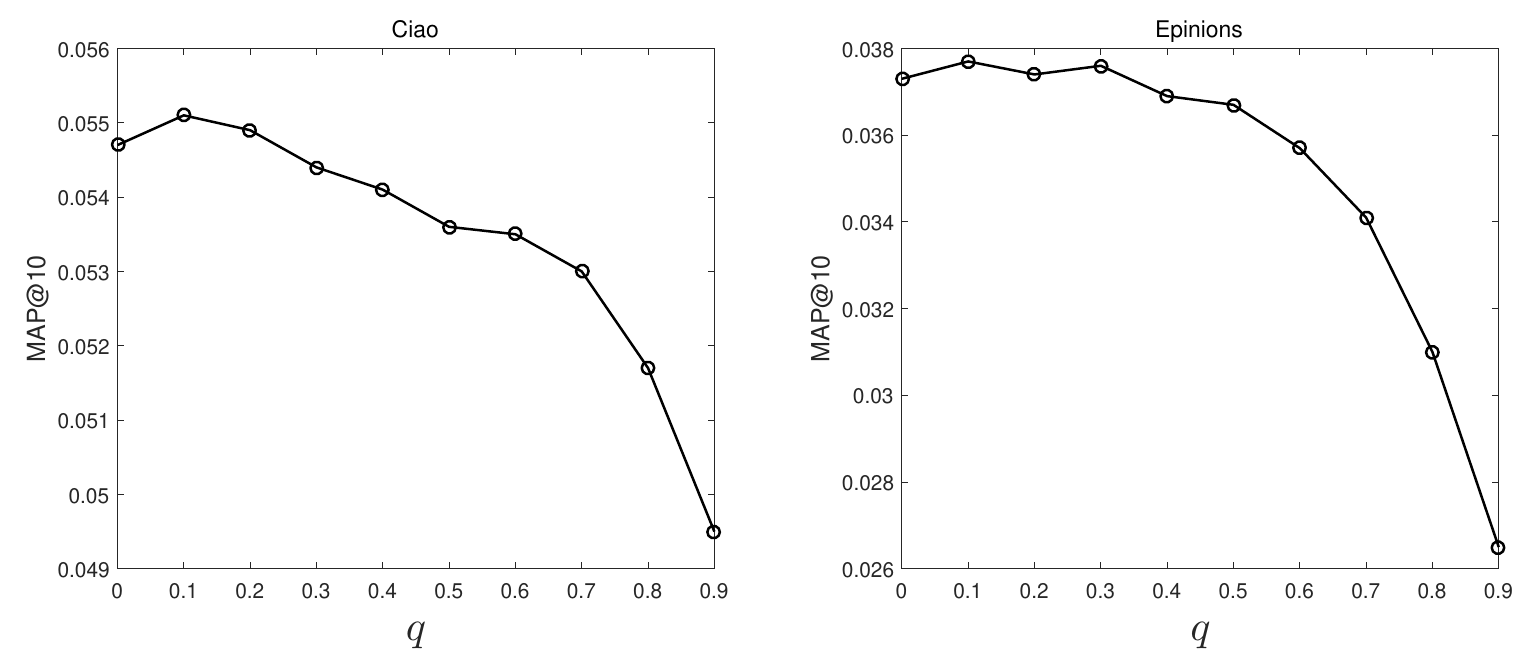}
		\caption{The impact of dropout probability $ q $.}
		\label{fig:dropout}
	\end{figure}
	
	To study the influence of $ q $ in CoDAE, we conduct a series of experiments to show the results with different values of $ q $. The parameter $ q $ are used control the dropout probability in Equation (\ref{eq:corr}). A larger value of $ q $ indicates the entries of input vectors are more likely to be dropout by replacing original values with zeros. If $ q=0 $, this dropout method makes no impact on the entries of input vectors.
	
	As demonstrated in Figure \ref{fig:dropout}, with the value of $ q $ increases, the performance of CoDAE grows to a particular point and then drops down for both datasets of Ciao and Epinions. Especially, the CoDAE model achieves best performance with $ q=0.1 $. Moreover, the performance of CoDAE with $ q=0 $ are slightly worse than that with $ q=0.1 $ for these two datasets. This is because of that the inputs of user ratings and social networks are facing critical problem, which makes it drops too much useful information by dropout method with a large value of $ q=0 $.

	\subsection{Impact of dimension $ k $}
	
	In this subsection, we study the impact of latent dimension $ k $ on Ciao and Epinions datasets and report the results in Table~\ref{tab:dimensions}. As demonstrated in table \ref{tab:dim}, we can see that the performance of CoDAE is increasing with $ k $ getting larger on both datasets. We can also see that the improvement degree is getting smaller with $ k $ grows or even getting worse due to overfitting, such as the result on Ciao and Epinions datasets with $ k=200 $. 
	
	\begin{table}[!h]
		\footnotesize
		\begin{center}
			\caption{Impact of dimension $ k $}\label{tab:dimensions}
			\begin{tabular}{l|c|c|c|c|c|c}
				\hline
				$ k $ &          5 &         10 &         20 &         50 &        100 &        200 \\
				\hline
				Ciao &     0.0505 &     0.0523 &     0.0528 &     0.0529 &     \textbf{0.0547} &     0.0526 \\
				\hline
				Epinions &     0.0235 &     0.0272 &     0.0278 &     0.0294 &     \textbf{0.0317} &     0.0313 \\
				\hline
			\end{tabular}  
			\label{tab:dim}
		\end{center}
	\end{table}
	
	\section{Conclusions}
	In this paper, we propose a novel top-\textit{n} recommendation algorithm CoDAE by modeling users with roles of raters, truster and trustee. Especially, we propose a tightly coupled structure CoDAE to learn compact and high-level influence from rating and trust data for each user. Moreover, this structure contains three DAE networks, the mid-layers of which are tightly interconnected through a novel related regularization term. This model is very flexible and easily be extended for other kinds of auxiliary information or other applications. We further conduct comprehensive experiments to compare the CoDAE model with other state-of-the-art algorithms. We also study the influence of hyper-parameters in CoDAE model.
	
	In the further, we intend to develop the proposed model for at least three potential directions but not limited. First, to further overcome the sparse problem of rating and trust data, it is necessary to introduce more information for recommender systems, such as images or videos. We intend to make use of the recent techniques~\cite{11} to improve accuracy for recommender systems. Second, distinct from that in computer vision domain, the rating data used in recommender systems are very sparse. This fact makes it difficult make full use of GPU power by most existing deep learning frameworks, such as TensorFlow or pyTorch. This problem appeals for more attentions to introduce multi-core CPU or many-core GPU power~\cite{45,46} for recommender systems. Third, the proposed CoDAE is a very flexible model to be extended for other kinds of applications, such as CAD/CAM~\cite{47,48}, social computing~\cite{49,50} and intelligent computing~\cite{51,52}.
	
	\section*{Acknowledgments}	
	This research has been supported by the National Science Foundation of China (Grant No.61472289) and the National Key Research and Development Project (Grant No.2016YFC0106305).

\end{document}